\documentclass[twocolumn,showpacs,preprintnumbers,amsmath,amssymb]{revtex4}

\usepackage{graphicx}
\usepackage{dcolumn}
\usepackage{bm}
\usepackage[usenames]{color}

\begin{document}


\title{{\it Ab initio} investigation of ${\rm Fe^{2+}}$/${\rm Fe^{3+}}$ dimerization and ferroelectricity in multiferroic magnetite: role of electronic correlations}

\author{Tetsuya Fukushima}
\author{Kunihiko Yamauchi}
\author{Silvia Picozzi}%
\email{silvia.picozzi@aquila.infn.it}
\affiliation{%
Consiglio Nazionale delle Ricerche - Instituto Nazionale di Fisica della Materia (CNR-INFM), CASTI Regional Lab., 67100 L'Aquila, Italy\\
}%

\date{\today}
\begin{abstract}
Based on {\it ab initio} density functional theory, we have investigated a microscopic mechanism that leads to ${\rm Fe^{2+}_{B}}/{\rm Fe^{3+}_{B}}$ dimerization and consequent ferroelectricity in charge ordered ${\rm Fe_{3}O_{4}}$ with $P2$ symmetry. In addition to the simple inter-site Coulomb repulsion, quantum hybridization effects  are invoked to explain the ${\rm Fe^{2+}_{B}}/{\rm Fe^{3+}_{B}}$ bond dimerization. 
Our results, based on the  generalized gradient approximation + Hubbard $U$ (GGA+$U$) method, indicate that noncentrosymmetric $P2$ magnetite  shows a finite and sizeable ferroelectric polarization
along the $b$ crystalline axis. From the $U$ dependence of polarization, we conclude that the origin of ferroelectricity in $P2$ ${\rm Fe_{3}O_{4}}$ lies in the recently proposed ``intermediate site/bond-centered charge ordering".
\end{abstract}

\pacs{75.50.Gg, 77.80.-e}
\maketitle

\section{Introduction}

Advanced multifunctional materials
showing coexisting and spontaneous long-range dipolar and magnetic orders are 
nowadays referred to as {\em multiferroics} \cite{review}; they combine extremely fascinating physics phenomena with  a huge potential for technological applications. Recently, large interests have been devoted to {\em improper} multiferroics \cite{review,jpcm}: here, ferroelectricity is driven by unconventional mechanisms, such as the loss of centrosymmetry due to the occurrence of long-range {\em spin} or {\em charge} order (at variance with more conventional ferroelectrics in which the ionic degrees of freedom play a prominent role in inducing permanent dipole moments).

Magnetite, ${\rm Fe_{3}O_{4}}$ (formally ${\rm Fe_{A}^{3+}[Fe^{2.5+}Fe^{2.5+}]_{B}O_{4}^{2-}}$), crystallizes in the inverted cubic spinel structure $Fd\bar{3}m$ and shows a metallic behavior at room temperature. ${\rm Fe_{A}}$ and ${\rm Fe_{B}}$ ion sites are coordinated to O ions in the inverted spinel structure ${\rm Fe_{3}O_{4}}$, {\it i.e.,} the tetrahedrally surrounded Fe sites are occupied by ${\rm Fe_{A}}$ atoms, whereas the octahedrally coordinated Fe sites are occupied by ${\rm Fe_{B}}$ atoms. The latter form a pyrochlore lattice, which consists of a network of corner sharing tetrahedra. Ferrimagnetism is stable in ${\rm Fe_{3}O_{4}}$, with magnetic moments of ${\rm Fe_{A}}$ sites antiparallel to those of ${\rm Fe_{B}}$ sites. ${\rm Fe_{3}O_{4}}$ undergoes a first order metal-insulator transition (called Verwey transition) at around 120 K, where the resistivity of ${\rm Fe_{3}O_{4}}$ increases by two orders of magnitude. \cite{Verwey1} This phase transition accompanies a change in the crystal structure of ${\rm Fe_{3}O_{4}}$, from cubic to monoclinic symmetry. Verwey proposed that the metal-insulator transition originates from a charge ordering at the ${\rm Fe_{B}}$ sites (${\rm Fe_{A}^{3+}[Fe^{2+}Fe^{3+}]_{B}O_{4}^{2-}}$), similar to strongly correlated hole-doped manganites. \cite{Verwey2} A full charge localization with ideal ${\rm Fe^{2+}_{B}}$ and ${\rm Fe^{3+}_{B}}$ does not occur, since the charge disproportionation is much smaller than the ideal 3+/4+ ionic picture, due to strong covalency effects. Anderson pointed out that, when putting two ${\rm Fe^{2+}}$ and two ${\rm Fe^{3+}}$ on each ${\rm Fe_{B}}$ tetrahedron, the number of ${\rm Fe^{2+}}$ and ${\rm Fe^{3+}}$ pairs is maximized and this gives rise to the lowest energy from the Coulomb repulsion point of view. \cite{Anderson} However, 
the Anderson criterion for low temperature ${\rm Fe_{3}O_{4}}$ are inconsistent with recent experiment results. \cite{Garcia,Wright1} As such, the charge ordering occurring at the Verwey transition in ${\rm Fe_{3}O_{4}}$ is still under discussion.

Ferroelectricity has been recently observed at low temperatures in ${\rm Fe_{3}O_{4}}$ and attracted great scientific interests. \cite{Medrano,Alexe} Khomskii {\it et al.} have proposed that  ${\rm Fe_{3}O_{4}}$ might show ferroelectricity by means of an intermediate site/bond-centered charge ordering, being one of the first multiferroic materials. \cite{Khomskii} 
We investigated ferroelectricity in ${\rm Fe_{3}O_{4}}$ by {\it ab initio} calculations, as showing that magnetite has non-centrosymmetric $Cc$ symmetry structure in the ground state and the origin of ferroelectricity in ${\rm Fe_{3}O_{4}}$ is not the site/bond-centered charge ordering, but a ``charge shift'' between specific Fe$_{\rm B}$ site. \cite{Yamauchi1}

In contrast to our previous study, in this paper we focus on noncentrosymmetric $P2$ (No.3) structure in order to realize the picture proposed by Khomskii, where the ferroelectric polarization arises from the Fe$^{2+}_{\rm B}$/Fe$^{3+}_{\rm B}$ dimerization.
We elucidate three physical properties in ${\rm Fe_{3}O_{4}}$ based on {\it ab initio} density functional theory. The first point is a mechanism for ${\rm Fe^{2+}_{B}}/{\rm Fe^{3+}_{B}}$ dimerization along the monoclinic $b$ direction. Our results indicate that  the hybridization of electronic orbitals between ${\rm Fe^{2+}_{B}}$ sites ({\it i.e.} a quantum mechanical effect) is as important as the classical inter-site Coulomb interactions for the ${\rm Fe^{2+}_{B}}/{\rm Fe^{3+}_{B}}$ bond dimerization. The second point is related to ferroelectricity in $P2$ ${\rm Fe_{3}O_{4}}$ as induced by intermediate site/bond-centered charge ordering proposed by Khomskii {\it et. al}. \cite{Khomskii} Ferroelectric polarizations (calculated via the Berry-phase approach) are compared to those obtained in Ref.\cite{Yamauchi1} for $Cc$ magnetite. Finally, we investigate the stability of antiferroelectric $P2/c$ (No.13) and ferroelectric $P2$ ${\rm Fe_{3}O_{4}}$ vs electronic correlation effects.

\section{Electric polarization induced by charge ordering: basic concepts}
\begin{figure} [t]
\begin{center}
\includegraphics[width=8.5cm,clip]{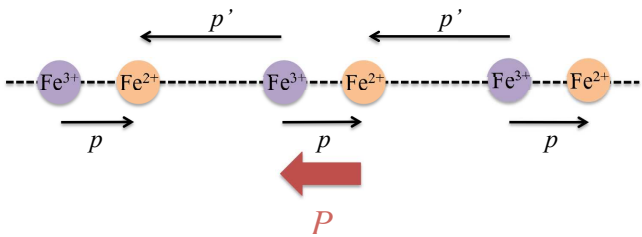}
\caption{Illustration of a net polarization (red arrow) by the intermediate site-centered and bond-centered charge ordering in ${\rm Fe_{3}O_{4}}$. The ${\rm Fe^{2+}_{B}}$ and ${\rm Fe^{3+}_{B}}$ charge ordering and an alternation of {\it short} and{ \it long} bond lengths between ${\rm Fe^{2+}_{B}}$ and ${\rm Fe^{3+}_{B}}$ sites in low temperature magnetite are shown.}
\label{dimmer}
\end{center}
\end{figure}
Khomskii {\it et al}. have discussed how charge ordering could induce ferroelectricity in magnetic systems. \cite{Khomskii}  According to their idea, when either site-centered or bond-centered charge orderings happen individually, a system does not have a net dipole moment: however, if the site-centered and bond-centered charge orderings occur simultaneously in a system, the inversion symmetry is broken so that the system shows a net dipole moment, therefore becoming ferroelectric ({\it cf}. Fig.\ref{dimmer}). \cite{Khomskii} Based on this scenario, Khomskii {\it et al.} have proposed that ${\rm Fe_{3}O_{4}}$ could become ferroelectric, with the monoclinic $b$ axis as polarization direction. Actually, experimental results indicate that the situation in  ${\rm Fe_{3}O_{4}}$ is similar to the intermediate site/bond-centered charge ordering: Wright {\it et al.} \cite{Wright2} have investigated the crystal structure of ${\rm Fe_{3}O_{4}}$ at low temperature by using high-resolution neutron powder-diffraction and X-ray powder-diffraction measurements. According to Ref.\cite{Wright2}, the distances between charge ordered ${\rm Fe^{2+}_{B}}$/${\rm Fe^{3+}_{B}}$ sites along  $b$ are strongly modulated, from 2.86 ${\rm \AA}$ to 3.05 ${\rm \AA}$. So, in addition to the site-centered charge ordering of ${\rm Fe^{2+}_{B}}$ and ${\rm Fe^{3+}_{B}}$ sites, there is a sequence of {\it short} and {\it long} ${\rm Fe_{B}}$-${\rm Fe_{B}}$ bond lengths along  $b$ ({\em i.e.} bond dimerization process). This situation indeed corresponds to the site/bond-centered charge ordering proposed by Khomskii  {\it et al} \cite{Khomskii}., the so called ``Zener polaron''.

\begin{figure} [t]
\begin{center}
\includegraphics[width=8.5cm,clip]{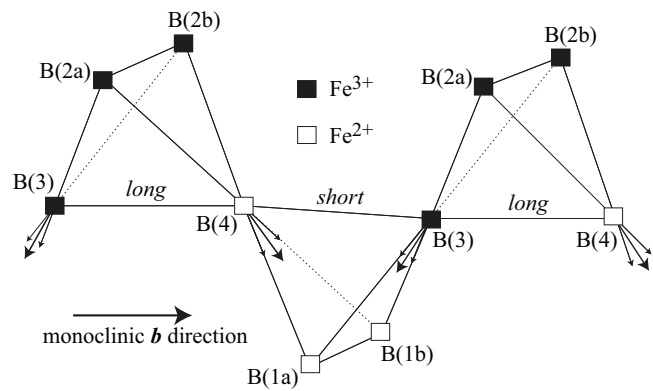}
\caption{Schematics of the corner sharing tetrahedra formed by ${\rm Fe_{B}}$ sites in $P2/c$ ${\rm Fe_{3}O_{4}}$. The labels of ${\rm Fe_{B}}$ site are consistent with Ref.\onlinecite{Wright2}. Black and white squares indicates ${\rm Fe^{3+}_{B}}$ and ${\rm Fe^{2+}_{B}}$ sites, respectively.}
\label{tetra_inter}
\end{center}
\end{figure}

Generally, the origin of bond dimerization in other multiferroic materials is the spin configuration,  {\it e.g.,} in ${\rm HoMn_{2}O_{5}}$ the bond lengths between antiparallel (parallel) Mn ions are expanded (contracted) so as to optimize the double exchange mechanism. \cite{gianluca} There is, however, no correlation between the dimerization process and the spin configuration in ${\rm Fe_{3}O_{4}}$, since all the spin directions of ${\rm Fe_{B}}$ sites are the same. Wright {\it et al.} have proposed the bond length modulation to be due to the simple classical inter-site Coulomb repulsion between charge ordered ${\rm Fe^{2+}_{B}}$ and ${\rm Fe^{3+}_{B}}$ sites, {\it i.e.,} below the Verwey temperature the corner sharing  ${\rm Fe_{B}}$ tetrahedra are constructed by three ${\rm Fe^{3+}_{B}}$ and one ${\rm Fe^{2+}_{B}}$ ions or by three ${\rm Fe^{2+}_{B}}$ and one ${\rm Fe^{3+}_{B}}$ ions (3-1 charge order), at variance with the Anderson criterion in which the configurations of two ${\rm Fe^{2+}_{B}}$ and two ${\rm Fe^{3+}_{B}}$ ions (2-2 charge order) are most stable \cite{Anderson}, as shown in Fig.\ref{tetra_inter}. It can be easily understood that a system can lower its energy, from  the {\em inter-site} Coulomb interaction point of view, if ${\rm Fe_{B}}$ ions on B(3) and B(4) sites move along the thin arrows in Fig.\ref{tetra_inter}. As a result, ${\rm Fe_{B}}$ ions on B(3) and B(4) sites are moved along the bold arrows in Fig.\ref{tetra_inter}, the net effect being a bond dimerization between B(3) and B(4) sites. However, as we'll show, an important role in the dimerization process in ${\rm Fe_{3}O_{4}}$ is played by the {\em on-site} Coulomb interaction of Fe atoms as well, and understanding the dimerization process in ${\rm Fe_{3}O_{4}}$ is not so trivial.

\section{Calculation method}\label{CM}
\begin{figure}[t]
\begin{center}
\includegraphics[width=8.5cm,clip]{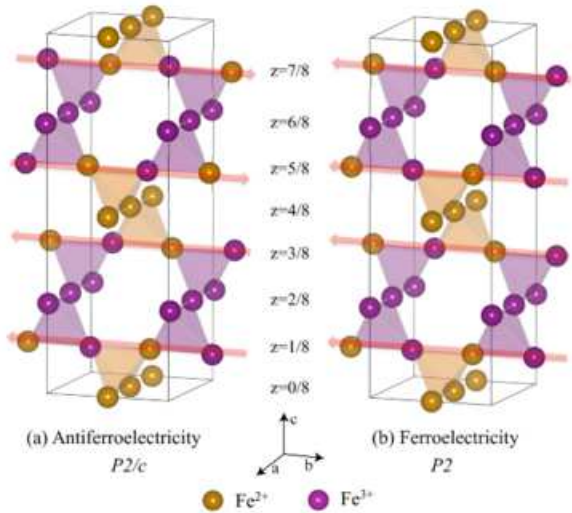}
\caption{Red arrows show local polarization directions in (a) antiferroelectric $P2/c$ and (b) ferroelectric $P2$ ${\rm Fe_{3}O_{4}}$. Purple and orange spheres indicate ${\rm Fe^{3+}_{B}}$ and ${\rm Fe^{2+}_{B}}$ ions, respectively. For clarity, ${\rm Fe_{A}}$ and O ions are not shown. Tetrahedra composed by three ${\rm Fe^{2+}_{B}}$ and one ${\rm Fe^{3+}_{B}}$ ions and three ${\rm Fe^{3+}_{B}}$ and one ${\rm Fe^{2+}_{B}}$ ions are highlighted by orange and purple planes, respectively.}
\label{p2c_p2}
\end{center}
\end{figure}
Electronic structure calculations and structural optimizations are performed by the "Vienna {\it Ab initio} Simulation Package" (VASP) and Projector Augmented Wave (PAW) pseudopotentials. \cite{Kresse} The Perdew-Becke-Erzenhof (PBE) of the generalized gradient approximation (GGA) is employed for the exchange-correlation potential. \cite{Perdew} The plane wave cutoff energies are 400 eV for Fe and O atoms. The GGA+$U$ calculations within Dudarev's approach are performed by applying a Hubbard-like potential for Fe-$d$ states only. \cite{Dudarev} Effective Coulomb energies $U=4.5$, $6.0$, and $8.0$ eV and an exchange parameter $J=0.89$ eV are used. \cite{Jeng1} To calculate the electronic structure of ferroelectric ${\rm Fe_{3}O_{4}}$, we consider an``artificial'' crystal structure with $P2$ symmetry, built by copying the lower half of the (experimentally suggested) monoclinic $P2/c$ antiferroelectric cell to the upper half, as shown in Fig.\ref{p2c_p2}. Note that there is no net polarization in antiferroelectric $P2/c$ magnetite, because the local polarizations in the upper half and lower half cancel each other. On the other hand, ferroelectric $P2$ ${\rm Fe_{3}O_{4}}$ can show a finite net polarization. Computational details for the electronic structure and structural optimization in ferroelectric $P2$ ${\rm Fe_{3}O_{4}}$ are the same as antiferroelectric $P2/c$. Both $P2/c$ and $P2$ contain 56 atoms  (24 Fe and 32 O ions) in $a_{\rm c}/\sqrt{2}{\times}a_{\rm c}/\sqrt{2}{\times}2a_{\rm c}$ unit cell, where $a_{\rm c}$ stands for the lattice constant of cubic $Fd\bar{3}m$ ${\rm Fe_{3}O_{4}}$. Lattice parameters are fixed to experimental values: $a=5.9341$ \AA, $b=5.9256$ \AA, $c=16.7524$ \AA, and $\beta=90.2000^{\circ}$. \cite{Zuo} In the structural optimization, a threshold on the atomic forces of 0.01 eV/${\rm \AA}$ is employed. Internal atomic coordinates are optimized in $P2/c$ and $P2$ ${\rm Fe_{3}O_{4}}$ starting from experimental Wyckoff parameters. \cite{Wright2} Fe-$3p^{6}3d^{6}4s^{2}$ and O-$2s^{2}2p^{4}$ electrons are treated as valence electrons. For Fe atoms, the ferrimagnetic configuration is considered: all ${\rm Fe_{A}}$ sites have up-spin and all ${\rm Fe_{B}}$ sites have down-spin. We neglected spin-orbit coupling, so that the direction of magnetization with respect to the crystal is not specified. For the $P2/c$ and $P2$ ${\rm Fe_{3}O_{4}}$, the $6{\times}6{\times}2$ Monkhost-Pack $k$-point grid in the Brillouin zone is used. The Berry phase approach \cite{Vanderbilt,Resta} within the PAW formalism and  the point charge model with nominal charges (${\rm Fe_{A}}$: 3+, ${\rm Fe^{2+}_{B}}$: 2+, ${\rm Fe^{3+}_{B}}: 3+$, and ${\rm O}$: 2-) are employed to calculate polarization. We integrate over six $k$-point strings parallel to the polarization direction (crystallographic $b$ direction) in the Berry phase approach.

\section{Results}
\subsection{${\rm Fe^{2+}}$/${\rm Fe^{3+}}$ bond dimerization in ${\rm Fe_{3}O_{4}}$}\label{dimerization}
\begin{figure} [t]
\begin{center}
\includegraphics[width=8.5cm,clip]{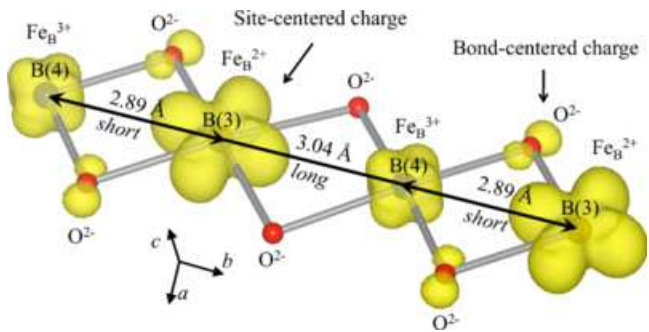}
\caption{Minority-spin charge densities of ${\rm Fe^{2+}_{B}}$, ${\rm Fe^{3+}_{B}}$, and ${\rm O^{2-}}$ sites in ferroelectric $P2$ ${\rm Fe_{3}O_{4}}$. A closeup of one of the z=1/8, 3/8, 5/8, and 7/8 planes in Fig.\ref{p2c_p2}(b). {\it Short} and {\it long} mean alternations of bond lengths between ${\rm Fe_{B}}$ sites due to the ${\rm Fe^{3+}}$/${\rm Fe^{2+}}$ bond dimerization along  $b$. Charge densities are calculated within 1.00 eV below the Fermi level.}
\label{bond_chg}
\end{center}
\end{figure}

First, we will show that the site/bond-centered charge ordering occurs in ${\rm Fe_{3}O_{4}}$, so that ferroelectricity can be expected. Figure \ref{bond_chg} illustrates charge densities (in a selected energy range, see caption) of ${\rm Fe^{2+}_{B}}$, ${\rm Fe^{3+}_{B}}$, and ${\rm O^{2-}}$ sites in ferroelectric $P2$ ${\rm Fe_{3}O_{4}}$ and an enlarged view of one of the z=1/8, 3/8, 5/8, and 7/8 planes in Fig.\ref{p2c_p2}(b). As discussed in a previous study \cite{Jeng2}, one electron occupies one of the spin down $t_{2g}$ orbitals ($xy$, $yz$, and $zx$) on ${\rm Fe^{2+}}$ sites ($d^{6}$ configuration, $t^{3}_{2g{\uparrow}}e^{2}_{g{\uparrow}}t^{1}_{2g{\downarrow}}$); therefore, charge densities on ${\rm Fe^{2+}}$ sites have a cross-like shape corresponding to $t_{2g}$ orbitals and extend to interstitial regions of anions ${\rm O^{2-}}$. The spin down $t_{2g}$ orbitals on ${\rm Fe^{3+}}$ sites ($d^{5}$ configuration, $t^{3}_{2g{\uparrow}}e^{2}_{g\uparrow}$) are pushed above the Fermi level.
The important point here is the charge density on the ${\rm O^{2-}}$ sites: charge is not present on ${\rm O^{2-}}$ sites in the {\it long} bonds ({\it cfr}. Fig.\ref{bond_chg}), whereas in the {\it short} bond sides there is an appreciable charge density with Oxygen $p$ character. The latter can be viewed as``bond-centered charge''. It can be easily seen that there is no overlap between $d$ orbitals on ${\rm Fe^{2+}_{B}}$ and ${\rm Fe^{3+}_{B}}$ sites and $p$ orbitals on ${\rm O^{2-}}$ sites in the {\it long} bond sides; however, the $d$ orbitals between ${\rm Fe^{2+}_{B}}$ and ${\rm Fe^{3+}_{B}}$ sites via $p$ orbitals on ${\rm O^{2-}}$ sites can hybridize on the {\it short} bond sides. Note that the hybridization between $d$ orbitals on ${\rm Fe^{2+}_{B}}$ sites and $p$ orbitals on ${\rm O^{2-}}$ sites is not perfectly $\pi$-like because of the small tilt of $p$ orbitals on ${\rm O^{2-}}$ sites, so that we can expect a somewhat larger hybridization with respect to the ideally undistorted atomic arrangement. As a result, due to orbital hybridization, {\it short} bond sides have higher electron density and {\it long} bond sides have lower electron density. The center of gravity of charges is therefore expected to move away from the ions and we infer from Fig.\ref{bond_chg} that, in addition to the {\em site}-centered charge ordering on ${\rm Fe_{B}}$ sites, even the {\em bond}-centered charge ordering proposed by Khomskii {\em et al.} is realized in ${\rm Fe_{3}O_{4}}$. \cite{Khomskii}
\begin{figure} [t]
\begin{center}
\includegraphics[width=8.5cm,clip]{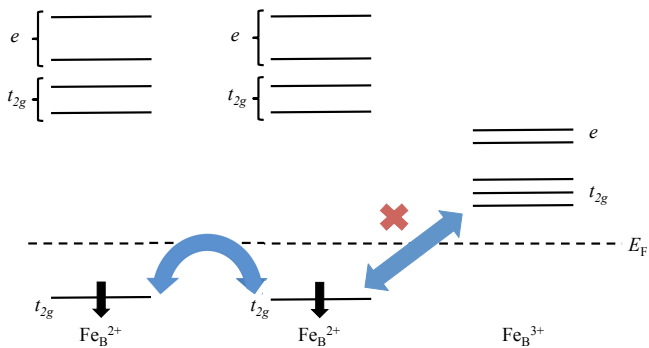}
\caption{Schematics of ``asymmetric hybridization'' between ${\rm Fe^{2+}_{B}}$ sites in the minority spin state. Thin black arrows denote the electron spin and thick blue arrows indicate electron hopping.}
\label{hybrid}
\end{center}
\end{figure}

Table \ref{table3} shows the dependence of ${\rm Fe_{B}}$-${\rm Fe_{B}}$ bond lengths on the on-site Coulomb interaction $U$ in the corner sharing tetrahedra and valence charges $q_{d}$ of ${\rm Fe_{B}}$ sites in ferroelectric $P2$ ${\rm Fe_{3}O_{4}}$ (the numbering of ${\rm Fe_{B}}$ sites corresponds to that of $P2/c$ structure in Fig.\ref{tetra_inter}). As shown in the upper part of Table \ref{table3} and Fig.\ref{tetra_inter}, there are alternations of $short$ and $long$ bond lengths between ${\rm Fe_{B}}$ sites, leading to the ${\rm Fe^{2+}}$/${\rm Fe^{3+}}$ bond dimerization along the $b$ axis, so that the corner sharing tetrahedra formed by ${\rm Fe_{B}}$ sites are strongly distorted from ideal regular tetrahedra. Table \ref{table3} also indicates that the distortions of the tetrahedra are suppressed upon increasing $U$. For example, along the $b$ direction, the bond lengths between ${\rm Fe_{B}}$(3) and ${\rm Fe_{B}}$(4) sites are modulated from 2.888 to 3.038 ${\rm \AA}$ for $U$=4.0 eV, from 2.919 to 3.006 ${\rm \AA}$ for $U$=6.0 eV, and from 2.934 to 2.992 ${\rm \AA}$ for $U$=8.0 eV, respectively (the bond length is modulated from 2.880 to 3.046 $\rm \AA$ in the experiment \cite{Wright1}).

Now, we focus on a mechanism of the ${\rm Fe^{2+}}$/${\rm Fe^{3+}}$ bond dimerization in ${\rm Fe_{3}O_{4}}$. Previous studies \cite{Wright2,Jeng1} suggested dimerization on ${\rm Fe^{2+}_{B}}$ and ${\rm Fe^{3+}_{B}}$ sites to originate from the inter-site Coulomb interaction between ${\rm Fe_{B}}$ sites. However, this is not the only factor and we propose that ``asymmetric hybridization'' plays a relevant role in this context ({\it cfr}. Fig. \ref{hybrid}). As mentioned above, one electron occupies the $t_{2g}$ down spin states at ${\rm Fe^{2+}_{B}}$ sites, whereas $t_{2g}$ and $e_{g}$ down spin states at ${\rm Fe^{3+}_{B}}$ sites are located above the Fermi level. In such conditions, ${\rm Fe^{2+}_{B}}$ electron orbitals can hybridize, so that the ${\rm Fe^{2+}_{B}}$-${\rm Fe^{2+}_{B}}$ bond lengths are shrinked. However, electron orbitals on ${\rm Fe^{2+}_{B}}$ and ${\rm Fe^{3+}_{B}}$ sites hybridize much less, due to the larger energy level difference between the $t_{2g}$ down spin states in ${\rm Fe^{2+}_{B}}$ and ${\rm Fe^{3+}_{B}}$ sites; therefore, shortening of bond lengths between ${\rm Fe^{2+}_{B}}$ and ${\rm Fe^{3+}_{B}}$ sites can not be expected by orbital hybridization. In this sense, we refer to the hybridization of electron orbitals on ${\rm Fe^{2+}_{B}}$ sites in ${\rm Fe_{3}O_{4}}$ as  ``asymmetric hybridization''. 

\begin{table*}[t]
\begin{center}
\begin{tabular}{ccccccc}
\hline \hline
$U$ (eV) &  B(1a)-B(1b) &  B(1a)-B(3) & B(1a)-B(4) & B(1b)-B(3) &  B(1b)-B(4) & B(2a)-B(2b) \\ \hline
4.5            &  2.9671         &  2.9215        & 2.9129       & 2.9118       &  2.9046       &  2.9671          \\
6.0            &  2.9671         &  2.9169        & 2.9366       & 2.9067       &  2.9289       &  2.9671          \\
8.0            &  2.9671         &  2.9156        & 2.9510       & 2.9058       &  2.9431       &  2.9671          \\ \hline
$U$ (eV) &  B(2a)-B(3)   &  B(2a)-B(4) & B(2b)-B(3) & B(2b)-B(4) &  B(3)-B(4) & B(4)-B(3) \\ \hline
4.5            &  3.0028         &  3.0195        & 3.0091       & 3.0247       &  {\bf 3.0379}       &  {\bf 2.8877}          \\
6.0            &  2.9958         &  3.0080        & 3.0017       & 3.0129       &  {\bf 3.0064}       &  {\bf 2.9193}          \\
8.0            &  2.9920         &  2.9977        & 2.9984       & 3.0031       &  {\bf 2.9923}       &  {\bf 2.9336}          \\ \hline \hline
                      $U$ (eV) & B(1a)   & B(1b)    & B(2a)   & B(2b)    & B(3)      & B(4)  \\ \hline
                       4.5          & 5.373 & 5.372 & 5.140 & 5.140 & 5.166 & 5.358       \\
                       6.0          & 5.398 & 5.397 & 5.106 & 5.106 & 5.116 & 5.401       \\
                       8.0          & 5.413 & 5.412 & 5.073 & 5.073 & 5.077 & 5.416       \\ \hline \hline
\end{tabular}
\caption{Upper and middle parts show the on-site Coulomb interaction $U$ dependences of ${\rm Fe_{B}}$-${\rm Fe_{B}}$ bond lengths (in ${\rm \AA}$) in the corner sharing tetrahedra. The lower part shows valence charges $q_{d}$ (in $e$) of ${\rm Fe_{B}}$ sites in $P2$ ${\rm Fe_{3}O_{4}}$. In the middle table, B(3)-B(4) and B(4)-B(3) indicate {\it long} and {\it short} bond sides, respectively.}
\label{table3}
\end{center}
\end{table*}

In order to prove the validity of the ``asymmetric hybridization'', we investigate the on-site Coulomb interaction dependence of valence charges $q_{d}$ of ${\rm Fe_{B}}$ ions (calculated within a Wigner-Seitz radius $r=1.00$$ {\rm \AA}$). As shown in the lower part of Table \ref{table3}, the ${\rm Fe^{2+}_{\rm B}}$/${\rm Fe^{3+}_{\rm B}}$ charge ordering occurs in ${\rm Fe_{3}O_{4}}$. This ${\rm Fe^{2+}_{\rm B}}$/${\rm Fe^{3+}_{\rm B}}$ charge ordering has been well known and investigated. \cite{Jeng1} We note that, when $U$ is set to zero, there is neither charge order nor bond dimerization of ${\rm Fe_{B}}$ sites, due to metallicity. The key point here is the $U$ dependence of charge separations between ${\rm Fe_{B}}$ sites. Our results indicate that, upon increasing $U$, the charge separation in ${\rm Fe_{B}}$ sites increases, due to the fact that  electrons are more localized for large $U$. Therefore, upon increasing the charge separations in ${\rm Fe_{B}}$ sites, one expects the inter-site Coulomb interactions to more strongly distort the ${\rm Fe_{B}}$ tetrahedra, implying a larger tendency towards bond dimerization between ${\rm Fe_{B}}$ sites. However, Table \ref{table3} shows that the dimerization on ${\rm Fe_{B}}$ sites decreases when increasing $U$. We infer that these results can not be explained by inter-site Coulomb interaction only. Indeed, from the ``asymmetric hybridization'' point of view, it can be expected that, due to the stronger localization of electrons for large $U$, the reduced hybridization leads to a weakening of dimerization on ${\rm Fe_{B}}$ sites (as confirmed by Table \ref{table3}). 

\subsection{Ferroelectricity in ${\rm Fe_{3}O_{4}}$}
From the discussion in the previous section, we can expect ferroelectric polarization along $b$ in $P2$ ${\rm Fe_{3}O_{4}}$ to occur by means of the site/bond-centered charge proposed by Khomskii {\it et. al}. \cite{Khomskii} Due to the presence of inversion symmetry, $P2/c$ ${\rm Fe_{3}O_{4}}$ doesn't have a net polarization; however, there is no inversion symmetry in $P2$ ${\rm Fe_{3}O_{4}}$, so the system is allowed to have finite polarization. We choose the antiferroelectric $P2/c$ ${\rm Fe_{3}O_{4}}$ as a centrosymmetric reference structure, {\it i.e.,} spontaneous polarization is obtained by polarization difference between ferroelectric $P2$ and antiferroelectric $P2/c$ ${\rm Fe_{3}O_{4}}$, ${\Delta}P=P_{\rm FE}-P_{\rm AFE}$. Here, polarization along $b$  in ${\rm Fe_{3}O_{4}}$ is calculated by two different methods: one is the Berry phase method, where the macroscopic polarization is obtained from the Wannier-center displacements, \cite{Vanderbilt} another is the simple ``point charge model'' with nominal charges on the Fe an O ions. Note that the polarization can only be determined up to an integer multiple of the polarization quantum \cite{Neaton}. In ferroelectric $P2$ ${\rm Fe_{3}O_{4}}$ structure, the polarization quantum $e\mbox{\boldmath $R$}/\Omega$ is 16.12 ${\rm {\mu}C/cm^{2}}$ along $b$ direction, where $e$ is the electron charge and $\mbox{\boldmath $R$}$ is the lattice vector along $b$.
\begin{figure} [t]
\begin{center}
\includegraphics[width=8.5cm,clip]{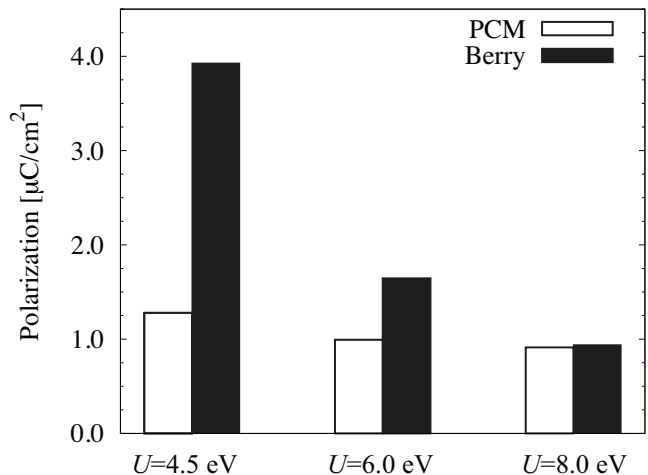}
\caption{Dependence of electric polarization $P^b$ in ferroelectric $P2$ ${\rm Fe_{3}O_{4}}$ on the on-site Coulomb interaction. Considered $U$ values are 4.5, 6.0, and 8.0 eV.  The exchange parameter $J$ is fixed to 0.89 eV. Polarizations calculated by point charge model and by the Berry phase method are shown.}
\label{polarization}
\end{center}
\end{figure}

Figure \ref{polarization} shows polarizations along  $b$ obtained by point charge model and by the Berry phase method. In the later approach, the internal atomic positions of ferroelectric $P2$ and antiferroelectric $P2/c$ ${\rm Fe_{3}O_{4}}$ are optimized in the conditions of $U$=4.5, 6.0, and 8.0 eV, and then the same $U$ values as for the crystal optimizations are consistently used for polarization calculations. At $U$=4.5 eV, the polarization calculated by the Berry phase method is much larger than that of the point charge model. This large difference is due to strong enhancement of dipole moments by the Wannier centers displaced from atomic centers. Such enhancement of polarization is of course not taken into consideration in the point charge model. Additionally, calculated polarizations decrease when increasing the $U$ value. The obtained results are consistent with the ${\rm Fe^{2+}_{\rm B}}$/${\rm Fe^{3+}_{\rm B}}$ bond dimerizations discussed in Sec.\ref{dimerization}. As increasing $U$, the ${\rm Fe^{2+}_{B}}$/${\rm Fe^{3+}_{B}}$ dimerizations  are suppressed, so it can be easily understood that if the bond dimerization between ${\rm Fe^{2+}_{B}}$ and ${\rm Fe^{3+}_{B}}$ sites disappears, polarization caused by the site/bond-centered charge ordering vanishes. Another important point here is the polarization with large $U$ values. As shown in Fig.\ref{polarization}, polarizations calculated by the point charge model and the Berry phase method for large $U$ approach the same value. This can be understood as follows: upon increasing $U$, electrons are more localized on Fe ions. Therefore, since the Wannier centers get closer to ${\rm Fe_{B}}$ sites and there is no enhancement of dipole moments by the displacements of the Wannier centers, the point charge model and the Berry phase method have similar values for large $U$.

We now comment on the difference between our present results and our previous calculations. \cite{Yamauchi1} We showed that ${\rm Fe_{3}O_{4}}$ has $Cc$ symmetry in the ground state and polarization direction is not along $b$ but in the monoclinic $ac$ mirror plane, $P_{\rm Berry}$=(-4.41, 0, 4.12) in ${\rm {\mu}C/cm^{2}}$ (for $U$=4.5 eV and $J$=0.89 eV). The difference between our present results and previous calculations originates from the microscopic mechanism of ferroelectricity: in ${\rm Fe_{3}O_{4}}$ with $Cc$ symmetry, the polarization is not driven by the site/bond-centered charge ordering but by "charge shift" at selected ${\rm Fe_{B}}$ sites. The bond dimerization on  ${\rm Fe^{2+}_{B}}$ and ${\rm Fe^{3+}_{B}}$ sites along $b$ in $Cc$ ${\rm Fe_{3}O_{4}}$ does indeed exist; however, dipole moments caused by the site/bond-centered charge ordering are cancelled between upper half and lower half of the unit cell, due to the $Cc$ symmetry. Therefore, the site/bond-centered charge ordering is not appropriate as a mechanism of ferroelectricity in $Cc$ magnetite.

\subsection{Stability of antiferroelectric $P2/c$ and ferroelectric $P2$ ${\rm Fe_{3}O_{4}}$}
\begin{figure} [t]
\begin{center}
\includegraphics[width=8.5cm,clip]{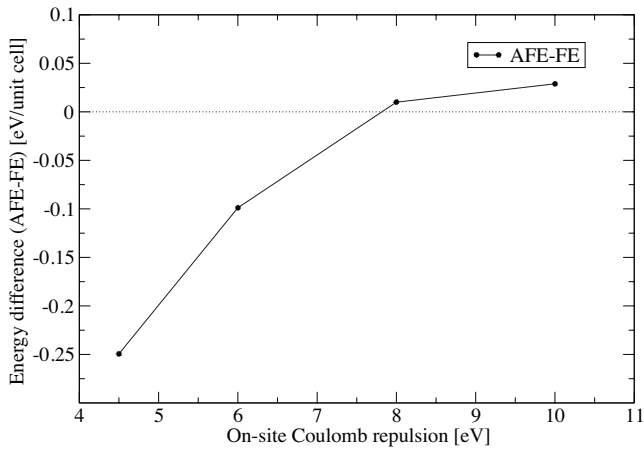}
\caption{On-site Coulomb interaction dependence of total energy difference per unit cell between antiferroelectric $P2/c$ (AFE) and ferroelectric $P2$ (FE) ${\rm Fe_{3}O_{4}}$.}
\label{ener_FE_AFE}
\end{center}
\end{figure}
Finally, we briefly discuss the stability of antiferroelectric $P2/c$ and ferroelectric $P2$ ${\rm Fe_{3}O_{4}}$. Figure \ref{ener_FE_AFE} shows the total energy differences per unit cell of antiferroelectric and ferroelectric ${\rm Fe_{3}O_{4}}$ (after structural optimizations) as a function of on-site Coulomb interaction $U$. The structural optimizations are performed separately for ferroelectric and antiferroelectric ${\rm Fe_{3}O_{4}}$ for $U$=4.5, 6.0. 8.0, and 10.0 eV. Incidentally, we note here that at $U$=4.5 eV, ${\rm Fe_{3}O_{4}}$ with $Cc$ symmetry is the ground state  \cite{Yamauchi1}. When the on-site Coulomb interaction is set to $U=4.5$ eV, the ferroelectric state is less stable than antiferroelectric state as shown in Fig.\ref{ener_FE_AFE}. However, the situation is changed for large $U$ values. The energy difference between antiferroelectric $P2/c$ and ferroelectric $P2$ ${\rm Fe_{3}O_{4}}$ decreases with increasing $U$ value; when $U$ reaches $\sim$8.0 eV the ferroelectric state becomes even more stable than the antiferroelectric state, paving the way to $P2$ magnetite as a possible multiferroic. 

\section{Summary}
Microscopic mechanisms leading to ${\rm Fe^{2+}_{B}}/{\rm Fe^{3+}_{B}}$ dimerization and ferroelectricity in ${\rm Fe_{3}O_{4}}$ have been investigated by {\it ab initio} density functional theory with Hubbard $U$ (GGA+$U$) method. From the on-site Coulomb interaction $U$ dependence of ${\rm Fe_{B}}$-${\rm Fe_{B}}$ bond lengths in corner sharing tetrahedra, we show that the "asymmetric hybridization" between ${\rm Fe^{2+}_{B}}$ sites plays an important role in ${\rm Fe^{2+}_{B}}/{\rm Fe^{3+}_{B}}$ bond dimerization, in addition to the simple inter-site Coulomb interaction. Additionally, based on the results of the $U$ dependence of polarizations and ${\rm Fe_{B}}$-${\rm Fe_{B}}$ bond lengths, we show that the noncentrosymmetric ${\rm Fe_{3}O_{4}}$ with $P2$ symmetry can become ferroelectric by means of the intermediate site/bond centered charge ordering proposed by Khomskii, where the ${\rm Fe^{2+}_{B}}/{\rm Fe^{3+}_{B}}$ site charge order and ${\rm Fe^{2+}_{B}}/{\rm Fe^{3+}_{B}}$ bond dimerization along the monoclinic $b$ direction simultaneously occurs. 
$P2$ ${\rm Fe_{3}O_{4}}$ can therefore be well considered as an {\it improper} multiferroic material with charge--order--induced ferroelectricity.

\section{Acknowledgments} 
The research leading to these results has received funding from the European Research Council under the European Community's 7th Framework Programme (FP7/2007-2013) / ERC grant agreement n. 203523. We thank Prof. Daniel Khomskii for useful discussions. The crystal structures and the charge states are plotted by using the software VESTA \cite{VESTA}.


\end{document}